\pdfoutput=1

\documentclass[11pt,a4paper]{article}
\usepackage{jinstpub}

\usepackage{booktabs}

\let\ifpdf\relax

\title{Monte Carlo study of the Coincidence Resolving Time of a liquid xenon PET scanner, using Cherenkov radiation}\author[a]{J.J.~Gomez-Cadenas,}
\author[a]{J.M.~Benlloch-Rodr\'iguez,}
\author[a,1]{P.~Ferrario}\note{Corresponding author.}
\emailAdd{paola.ferrario@ific.uv.es}
\affiliation[a]{Instituto de F\'isica Corpuscular (IFIC), CSIC \& Universitat de Val\`encia,\\
Calle Catedr\'atico Jos\'e Beltr\'an, 2, 46980 Paterna, Valencia, Spain}

\abstract{In this paper we use detailed Monte Carlo simulations to demonstrate that liquid xenon (LXe) can be used to build a Cherenkov-based TOF-PET, with an intrinsic coincidence resolving time (CRT) in the vicinity of 10 ps. This extraordinary performance is due to three facts: a) the abundant emission of Cherenkov photons by liquid xenon; b) the fact that LXe is transparent to Cherenkov light; and c) the fact that the fastest photons in LXe have wavelengths higher than 300 nm, therefore making it possible to separate the detection of scintillation and Cherenkov light.  The CRT in a Cherenkov LXe TOF-PET detector is, therefore, dominated by the resolution (time jitter) introduced by the photosensors and the electronics. However, we show that for sufficiently fast photosensors (e.g, an overall 40 ps jitter, which can be achieved by current micro-channel plate photomultipliers) the overall CRT varies between 30 and 55 ps, depending of the detection efficiency. This is still one order of magnitude better than commercial CRT devices and improves by a factor 3 the best CRT obtained with small laboratory prototypes. 
}


\begin{document}

\maketitle

\section{Introduction}

\subsection{Positron Emission Tomography}
Positron emission tomography (PET) is a functional imaging technique used to observe metabolic processes in the body. A PET apparatus detects pairs of gamma rays emitted indirectly by a positron-emitting radionuclide (tracer), which is introduced into the body on a biologically active molecule. The positron annihilates with an electron of the neighbouring atoms, producing two 511-keV photons with momenta on the same line (line of response, or LOR), but in opposite directions. A sensor system surrounding the patient detects the coincidence of the two gammas and constructs their  LOR. Crossing many LORs yields the image of the area where the radiotracer concentrates. 

The measurement of the time difference between the arrival of the two photons (time--of--flight, or TOF) improves the signal to noise ratio and the scanner sensitivity. The resolution in TOF (known as coincidence resolving time, or CRT) depends on: a) the physical properties of the radiator (e.g, the yield and the emission time of the photons); b) the detection efficiency; c) the time resolution (time jitter) introduced by the photosensors and the readout electronics.  

Current commercial PET scanners use inorganic scintillating crystals as detectors, such as LYSO (lutetium yttrium oxyorthosilicate), which enable TOF measurements, resulting in a CRT of $300-600$ ps FWHM \cite{gemini,vereos}. The most recent investigations performed in small laboratory systems, deploying detectors of a few tens of mm$^3$ volume, obtain a CRT of $\sim 80-120$ ps FWHM \cite{LysoCRT, LysoFBK}. 
 
\subsection{The PETALO concept}
Recently, a new scintillator detector, called PETALO\footnote{Positron Emission TOF Apparatus with Liquid xenOn} based on liquid xenon cells read out by silicon photomultipliers (SiPM)  has been proposed \cite{Gomez-Cadenas:2016mkq}. The excellent properties of liquid xenon as scintillator are clearly established in the literature \cite{Aprile:2009dv, Chepel:2012sj, Doke1, xemis, Nishikido2}. Specifically, LXe is attractive for two main reasons: a) it has a large scintillation yield (60 photons per keV of deposited energy) and b) its scintillation is a fast process, which can be parametrized as the sum of two exponentials with decay times of 2.2 and 27 ns. Furthermore, since LXe is a uniform, continuous medium, it permits the design of a homogeneous PET scanner (with minimal dead regions). At atmospheric pressure xenon liquefies at $\sim$ 161 K and thus cryogeny is relatively simple. On the other hand, operation at that temperature reduces the dark noise of SiPMs to negligible levels. In Ref.~\cite{Gomez-Cadenas:2016mkq} a Monte Carlo study was carried out to assess the TOF measurement performance of such a detector, using xenon scintillation light. If the scanner is equipped with (currently available) VUV-sensitive  SiPMs, an excellent CRT of 70  ps can be obtained.

In this paper, we present a Monte Carlo study of the CRT that can be reached in the PETALO detector using Cherenkov instead of scintillation light. In Sec.~\ref{sec.cher} we introduce the Cherenkov effect and its application to the PET technology. In Sec.~\ref{sec.MC} we describe the Geant4 simulation used for the current study and discuss the characteristics relevant for a CRT evaluation. In Sec.~\ref{sec.analysis} we discuss the performance of a PETALO scanner in CRT measurement using Cherenkov light. Conclusions are presented in Sec.~\ref{sec.concl}.
\section{The Cherenkov radiation}\label{sec.cher}

\subsection{The Cherenkov effect}\label{cher}
\begin{figure}[!bhtp]
	\centering
	\includegraphics[scale=0.2]{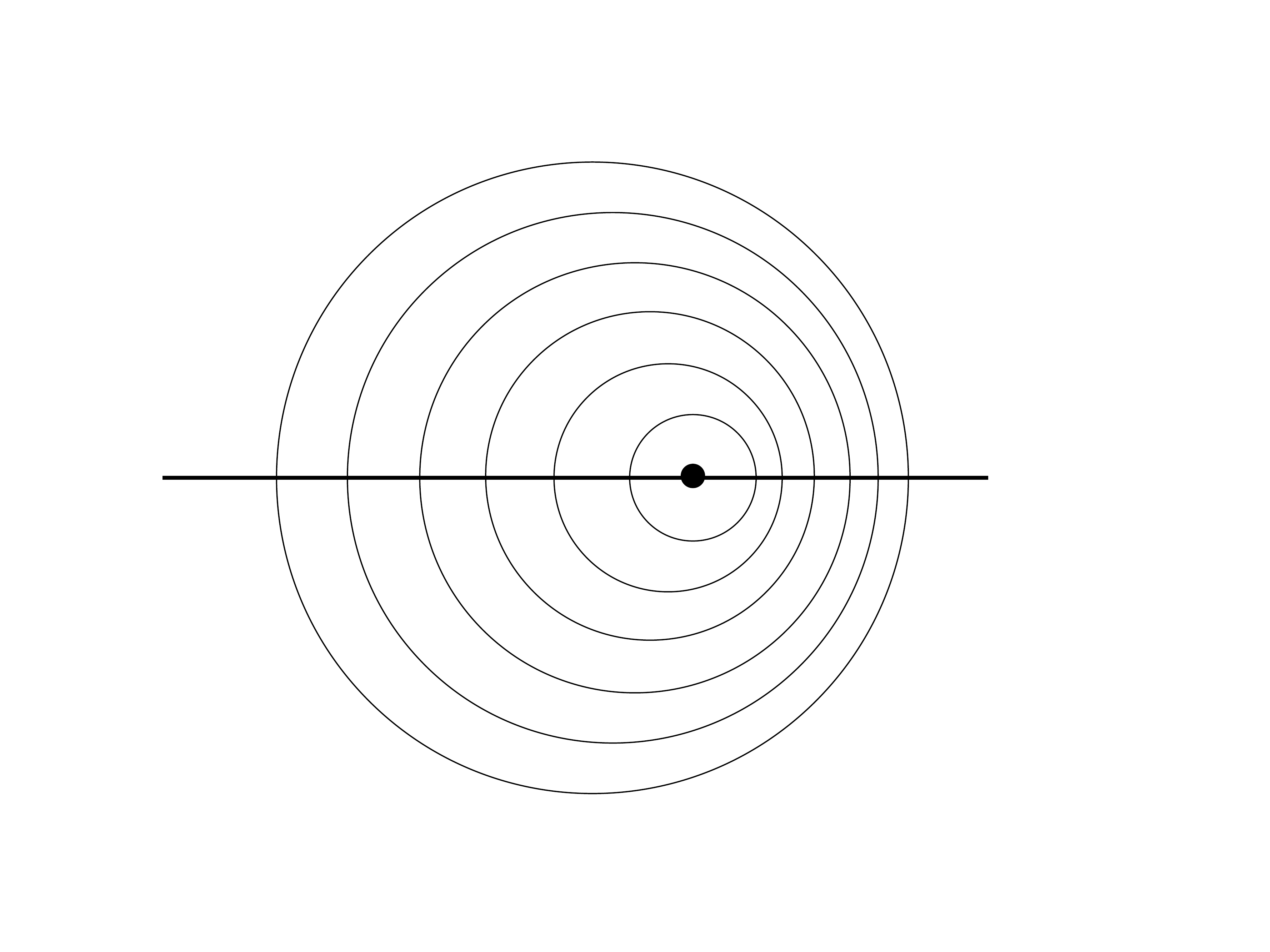}
	\includegraphics[scale=0.2]{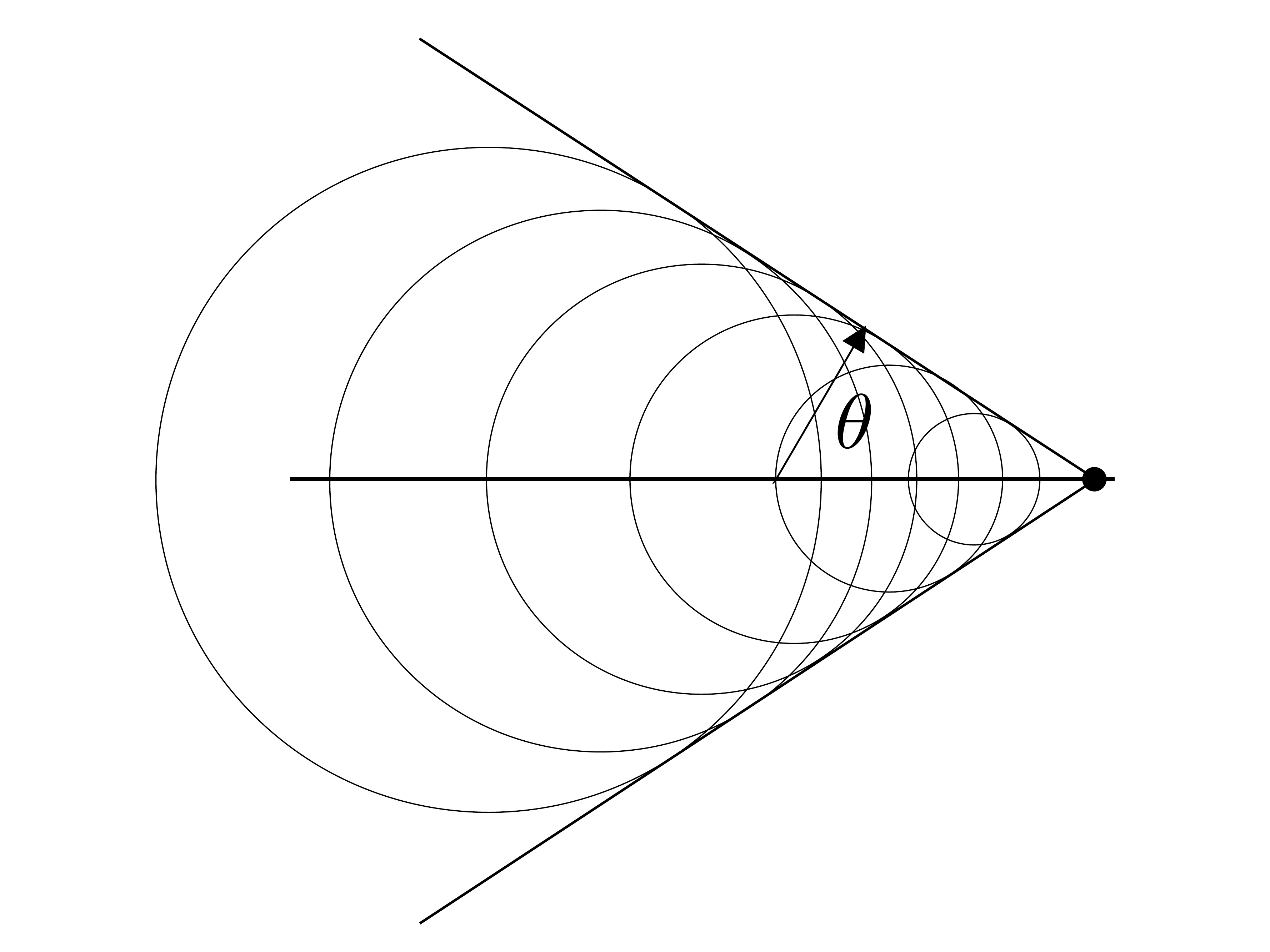}
	\caption{\label{fig.CherCone} Scheme of light emission when a particle travels at a speed lower (left), or higher (right) than that of light in the medium.}
\end{figure}

A charged particle propagating in a dielectric medium, at a speed higher than the speed of light in the medium, excites the surrounding molecules, which subsequently relax, emitting radiation. 
%
If a particle travels at a speed lower than that of light, the light emitted by the molecules of the medium at different points along its trajectory never interferes (see Fig.~\ref{fig.CherCone}\emph{-left}). However, if the particle speed is higher than that of light, the electromagnetic waves emitted at different points interfere constructively and emit a glow sufficiently intense as to be detected. The wave front propagates at an angle $\theta$ with the direction of the particle such that:
\begin{equation}
\cos\theta = \frac{v_{\textrm{light}}}{v}
\end{equation}
where $v_{\textrm{light}}$ is the speed of light in the medium and $v$ is the speed of the particle (see Fig.~\ref{fig.CherCone}\emph{-right}).
Cherenkov photons are emitted with wavelengths following the distribution:
\begin{equation}
\frac{\partial N}{\partial x\partial\lambda} = \frac{2\pi\alpha}{\lambda^2}\left(1-\frac{1}{\beta^2n^2(\lambda)}\right)
\label{eq.distrCher}
\end{equation}
where $\alpha$ is the fine-structure constant,  $\beta \equiv v/c$, $n$ is the refraction index of the medium, $\lambda$ is the photon wavelength and $x$ is the distance travelled by the charged particle. Eq.~\ref{eq.distrCher} does not diverge for high photon energy, because $n(\lambda)\rightarrow 1$ for short wavelengths (or high wave frequencies). Most Cherenkov radiation is emitted in the blue and ultraviolet range. Emission stops  when the speed of the particle drops below that of light in the medium, thus resulting in a threshold for light emission at a given wavelength of 
\begin{equation}
v>\frac{c}{n(\lambda)}
\label{eq.vthr}
\end{equation}
\subsection{Using Cherenkov light in PET scanners}
\begin{figure}[!bhtp]
	\centering
	\includegraphics[scale=0.35]{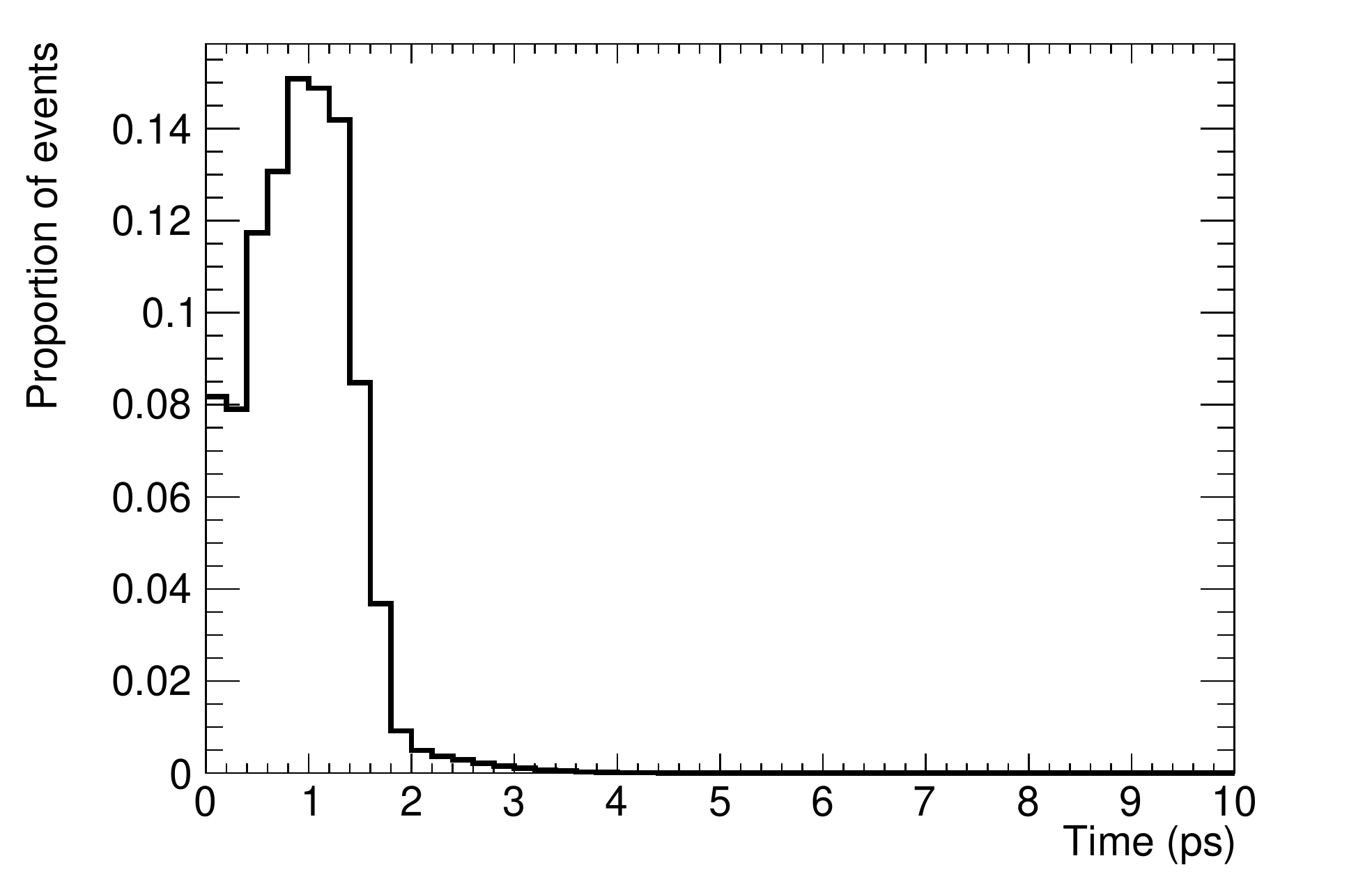}
	\includegraphics[scale=0.35]{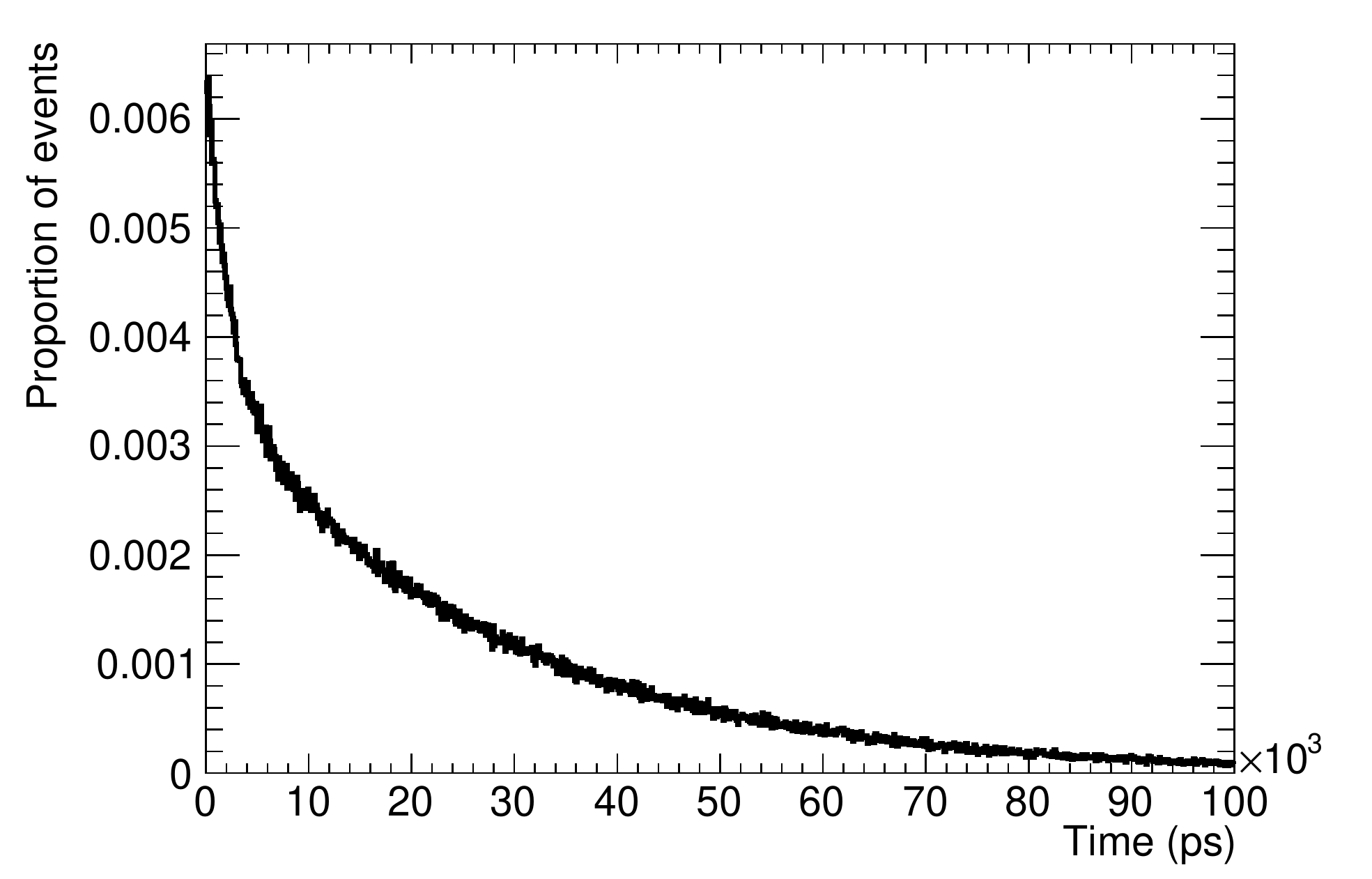}
	\caption{\label{fig.Times} Emission time of Cherenkov (left) and scintillation (right) photons in liquid xenon. Notice that the range of time axis for scintillation is $10^3$ times larger than that of Cherenkov radiation.}
\end{figure}

The promptness of Cherenkov light (few picoseconds, to be compared with tenths of nanoseconds for scintillation light, see Fig.~\ref{fig.Times}) is a very attractive feature for TOF applications of PET scanners, since it could lead to a dramatic improvement of the CRT, provided that: a) the yield of Cherenkov photons is sufficiently high and b) the noise introduced by photosensors and electronics is sufficiently low.  



%
In the last decade, the idea of using Cherenkov light for TOF-PET in scintillation crystals have been vigorously  pursued and numerous measurements as well as Monte Carlo studies have been carried out.  The main difficulty found is the very low efficiency of detecting two photons in coincidence, due to the high absorption of UV and blue light in conventional PET detectors, such as LYSO \cite{Brunner:2014}. The best results so far have been obtained with PbF$_2$ scintillator crystals read out by microchannel plate photomultipliers \cite{cherTOFKorpar}, which give a CRT of 71 and 95 ps FWHM in small setups (5 and 15 mm long respectively). 

Liquid xenon, on the other hand, is transparent to UV and blue light. Furthermore, scintillation and Cherenkov light separate naturally, as will be shown below. Last but not least, operation at moderate cryogenic temperatures may be an advantage (e.g, negligible dark count rate in sensors such as SiPMs \cite{dcr}). Therefore, liquid xenon appears to be an optimal candidate for TOF measurements using Cherenkov radiation.

\section{Monte Carlo simulation} \label{sec.MC}

\subsection{Description of the set-up} \label{setup.MC}
To study the performance of a Cherenkov radiation TOF-PET based on liquid xenon, we have simulated a two-cell set-up using the Geant4 toolkit with version number 10.01.p01 \cite{Agostinelli:2002hh,Allison:2006ve}. 
Our set-up is the same as the one described in Ref.~\cite{Gomez-Cadenas:2016mkq} and consists of two cells of 2.4 $\times$ 2.4 $\times$  5 cm$^3$ filled with liquid xenon, at a distance of 20 cm along the $z$ axis, on the opposite sides of a back-to-back 511 keV gamma source. The cells are instrumented in their entry and exit face with a dense array of 8 $\times$ 8 photosensors with an active area of 3 $\times$ 3 mm$^2$ and configurable photodetection efficiency. The photosensors are placed at a pitch of 3 mm, thus they cover the whole box face. The uninstrumented faces are covered by polytetrafluoroethylene (PTFE), which reflect optical photons according to a lambertian distribution with an efficiency of 97\%. This value for the reflectivity has been chosen following Ref.~\cite{Yamashita:2004} and references therein. A reflectivity higher than 99$\%$ is reported for the spectral range 350--1800 nm, and slightly lower values from 350 nm down to 200 nm. A simple extrapolation to 175 nm would give 95$\%$. For the scintillation spectrum of LXe (155--200 nm) values between 88$\%$ and 95$\%$ are found to produce a good fit of Monte Carlo to data, varying the absorption length of LXe to UV light from 1 m to infinity. We have chosen to use 97$\%$ for all wavelength as an average value and an absorption length $> 1$ m, which is virtually the same as infinity, given the dimensions of our cells. The physical properties used in the simulation which are relevant for the generation and propagation of optical photons are summarized in Table \ref{tab.LXeProp}, together with the main characteristics of the geometry.  
\begin{table*}[h!]\centering
\begin{tabular}{@{}l|l@{}}\toprule
Parameter & Value \\  \midrule
\multicolumn{2}{c}{Geometry}\\  \midrule
Cell dimensions & $2.4\times 2.4 \times 5$ cm$^3$ \\
Distance between cell entry faces & 20 cm \\
Sensor pitch & 3 mm\\
Number of sensors per face & 64 \\
\midrule
\multicolumn{2}{c}{Physics properties}\\  \midrule
LXe density & 2.98 g/cm$^3$ \cite{DensityLXe}\\
LXe attenuation length for 511 keV gammas & 4 cm \cite{Oberlack:2000cw} \\
LXe Rayleigh scattering length &  36.4 cm \cite{RayleighLXe} \\
Refraction index of sensor entrance window & 1.54 \\
Reflectivity of PTFE walls & 0.97 \\
 \end{tabular}
\caption{Summary of the Monte Carlo set-up specifications and the LXe relevant properties used in the simulation.}\label{tab.LXeProp}
\end{table*}
The coverage of the instrumented faces is assumed to be 100$\%$. We take into account the effect of a non-perfect coverage in a global detection efficiency (GDE), which encloses the probability of the photon to fall within the active area of a sensor and its photodetection efficiency.

\begin{figure}[!bhtp]
	\centering
	\includegraphics[scale=0.37]{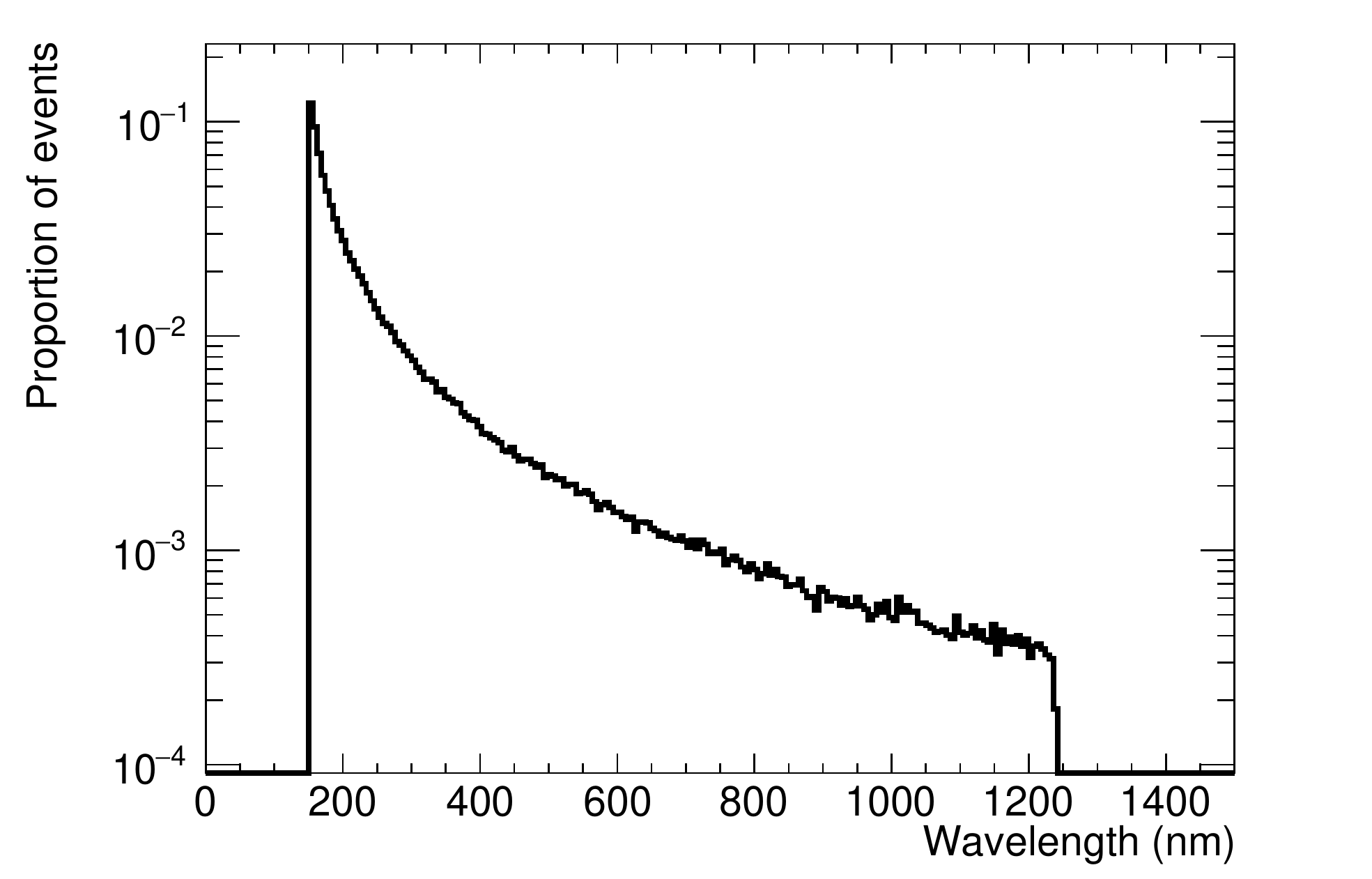}
	\includegraphics[scale=0.37]{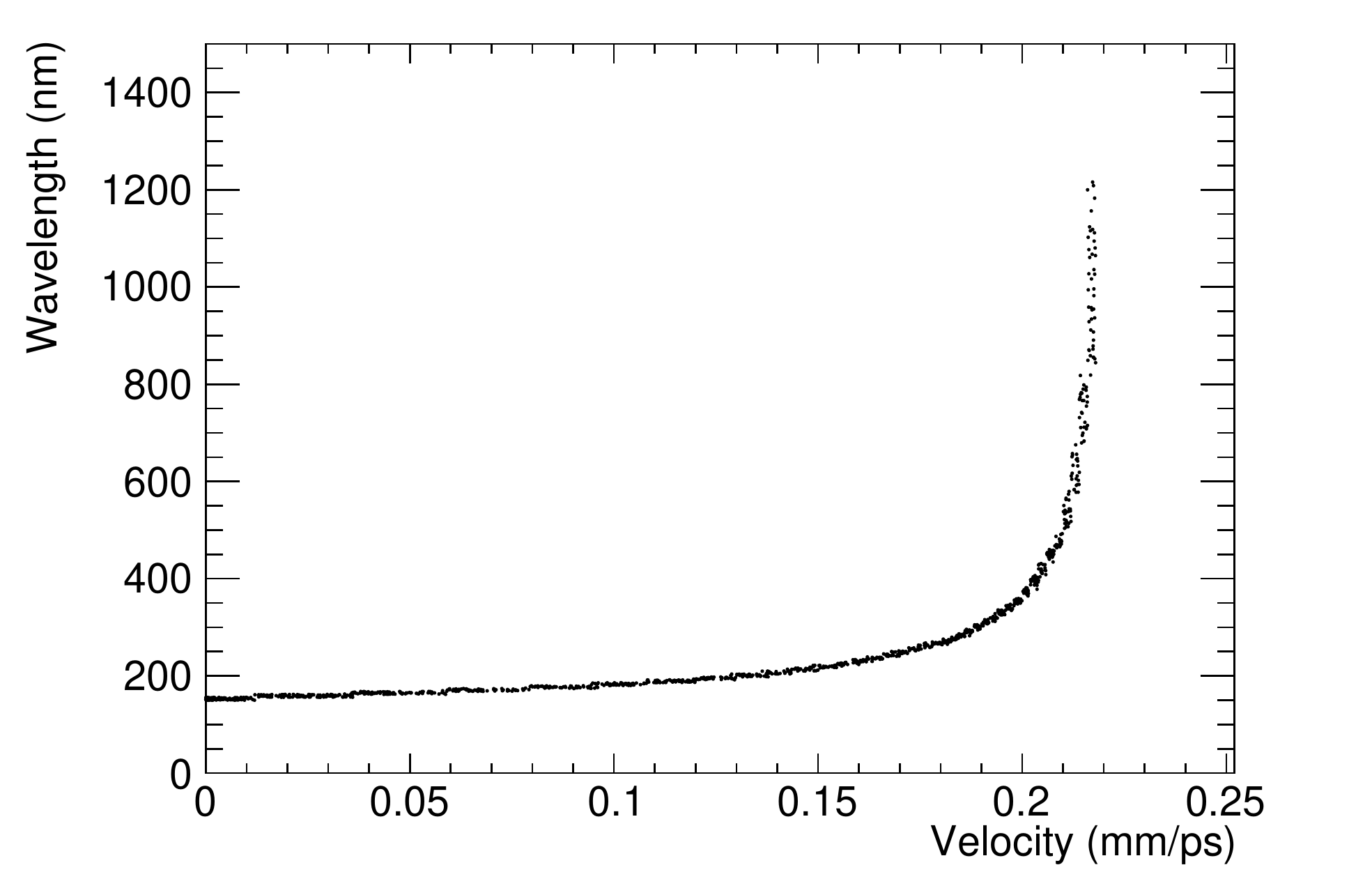}
	\caption{\label{fig.CherVel}(Left) Distribution of the Cherenkov radiation wavelength used in the simulation. (Right) Cherenkov radiation wavelength as a function of the group velocity of the photons. The plot shows that in the 350--600 nm range the photon velocity has very little variation.}
\end{figure}

Back-to-back 511 keV gammas are shot isotropically, at  $t=0$ ~from a vertex at equal distance from both boxes. The photons  can interact in the liquid xenon via photoelectric absorption or Compton scattering. In this study we focus on photoelectric events, only, in which all the available kinetic energy (511 keV) is deposited in one point. The Cherenkov radiation emitted by the electrons produced in such processes is simulated, with a wavelength distribution that follows Eq.~\ref{eq.distrCher} and is shown in Fig.~\ref{fig.CherVel}--\emph{left}. A cut is set at 155 nm, because no reliable measurements of the refraction index of liquid xenon exists for energies higher than $\sim$ 8 eV, which corresponds to $\sim$ 155 nm (see, for instance, Ref.~\cite{Baldini:2004td}). This assumption will not affect our results, since the fastest photons (which dominate the CRT, as will be demonstrated below) have much higher wavelengths (Fig.~\ref{fig.CherVel}--\emph{right}), mainly in the blue and near UV region. The upper cut at 1200 nm is conservative, being well beyond the typical sensitivity of the photosensors proposed.

\begin{figure}[!bhtp]
	\centering
	\includegraphics[scale=0.6]{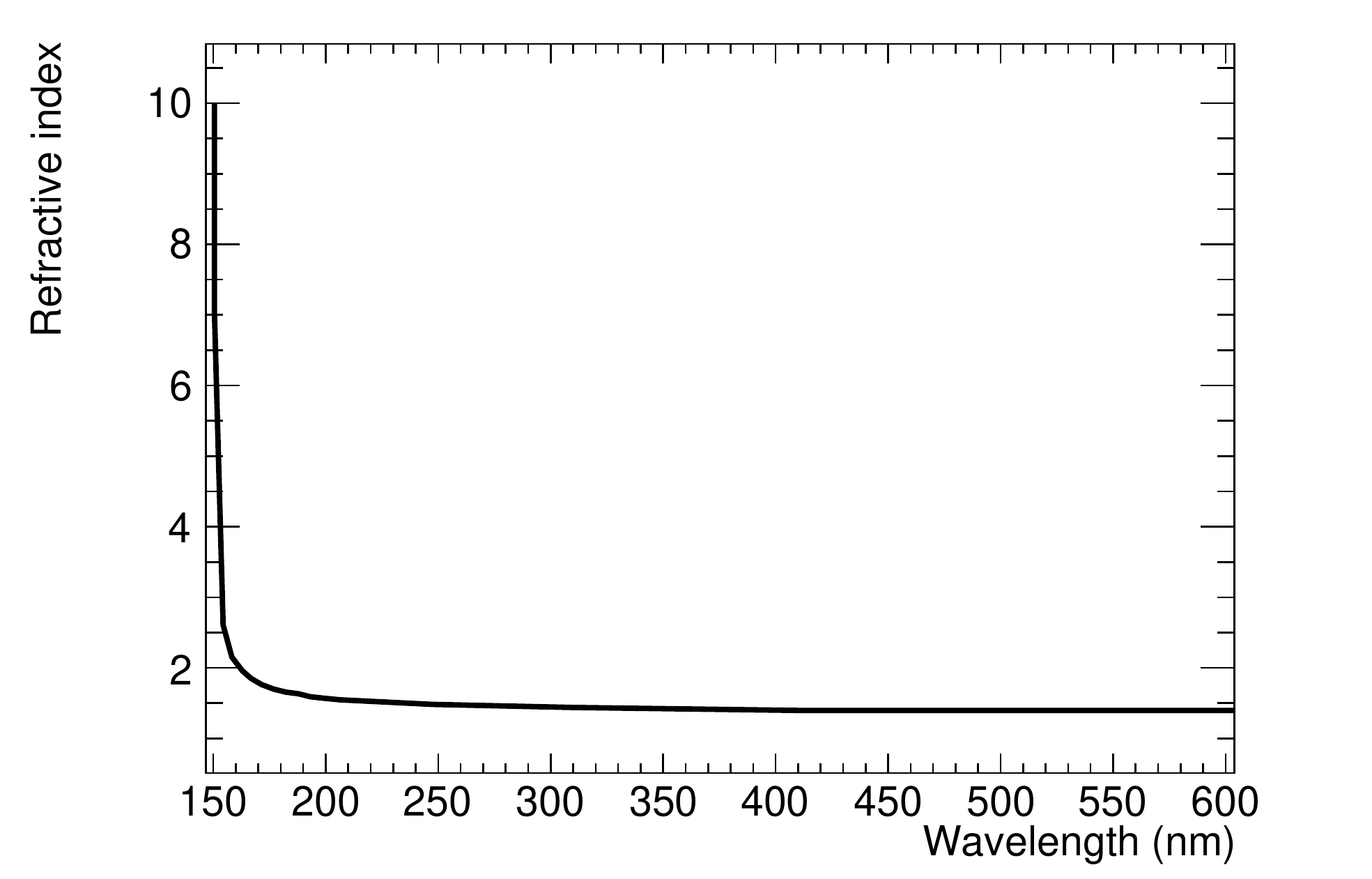}
	\caption{\label{fig.nlambda} LXe refraction index as a function of the wavelength of the optical photon, as results from the parametrization in Ref.~\cite{Baldini:2004td}. For wavelengths above 300 nm the refraction index is practically flat, and has a value of  around 1.4.}
\end{figure}

%
The Cherenkov photons are propagated inside the box  and eventually either reach a photosensor where they may produce a photoelectron, depending on the GDE, or are absorbed by the uninstrumented faces. The LXe refraction index dependence on the energy of photons is simulated according to the Lorentz-Lorenz equation \cite{Baldini:2004td}
\begin{equation}
 \frac{n^2 - 1}{ n^2 + 2}= - A (E) \cdot d_M
\end{equation}
where $n$ is the LXe refraction index, $d_M$ is the molar density  and $A(E)$ is the first refractivity viral coefficient:
\begin{equation}
A (E) = \sum_i^3 \frac{P_i}{E^2-E_i^2} 
\end{equation}
with $P_i \textrm{(eV$^2\cdot$ cm$^3$/mole)} = (71.23, 77.75, 1384.89)$ and $E_i \textrm{(eV)} = (8.4, 8.81, 13.2)$. This dependence is illustrated in Fig.~\ref{fig.nlambda} in terms of the wavelength.

\begin{figure}[!bthp]
	\centering
	\includegraphics[scale=0.6]{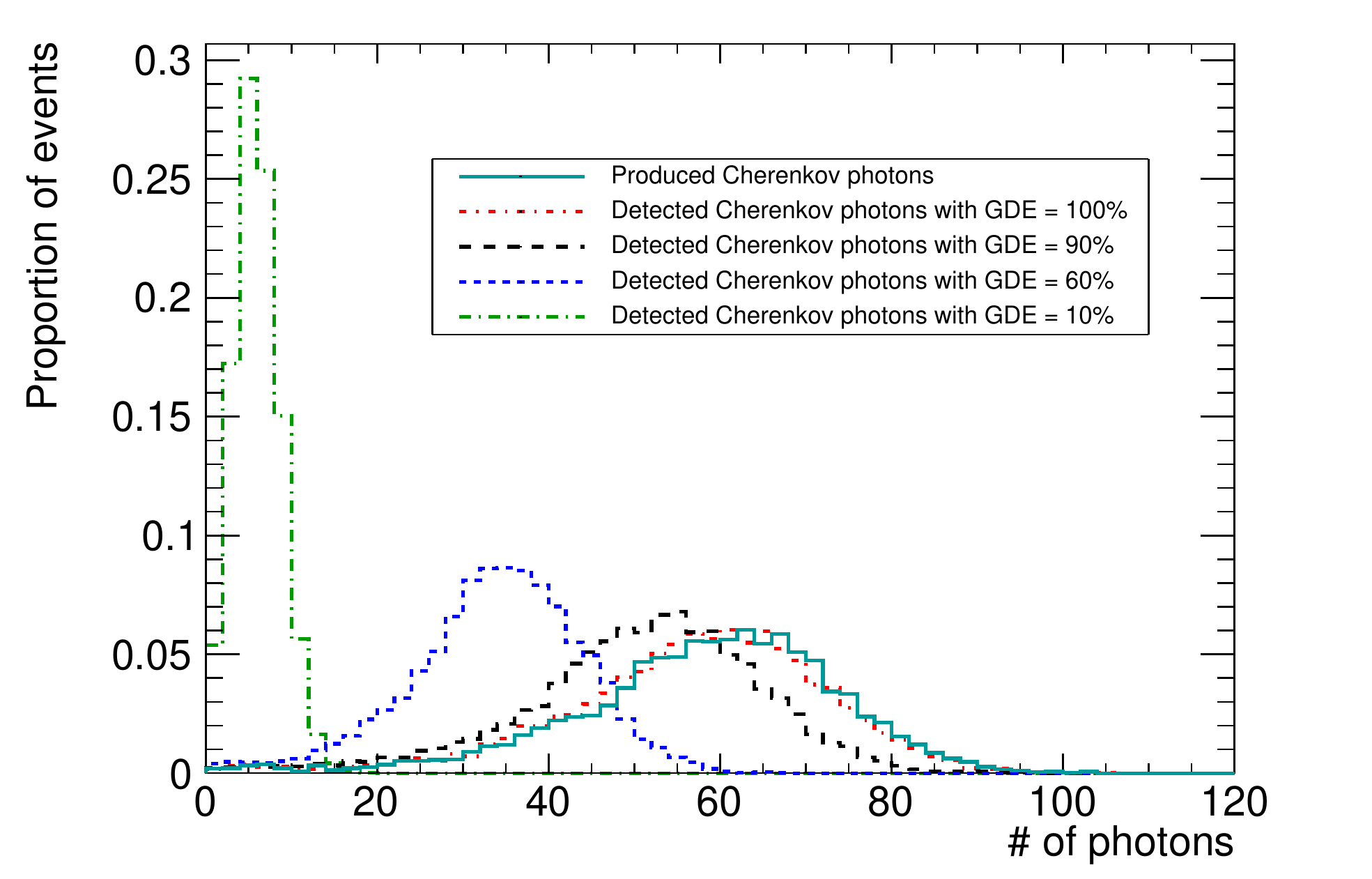}
	\caption{\label{fig.numbCher} Distribution of the number of Cherenkov photons produced and detected in one photoelectric interaction of a 511--keV gamma in our set-up, varying the GDE of the sensors between 10$\%$ and 100$\%$.}
\end{figure}

\begin{table*}[h!]\centering
\begin{tabular}{@{}c  | c | c | c@{}}\toprule
Bin range  & Refraction & Energy threshold & \# photons \\  
(nm) & index & (keV) &  for 511--keV e-\\  \midrule
155--250  & 2.58--1.46 & 43--189 & 41 \\
250--350  & 1.46--1.41 & 189--214 & 10 \\
350--450  & 1.41--1.39 & 214--224 &  5 \\
450--550  & 1.39 --1.38& 224--229 & 3 \\
550--650 & 1.38 & 229--232 & 2 \\
\midrule
 \end{tabular}
\caption{Properties of Cherenkov light simulation in liquid xenon.}\label{tab.CherProp}
\end{table*}
In Table \ref{tab.CherProp} the properties of Cherenkov radiation production in liquid xenon are summarized for the wavelengths of relevance in this study. The kinetic energy threshold $E_{\textrm{thr}}$ for electrons to produce Cherenkov radiation at a fixed wavelength $\lambda$ is calculated by
\begin{equation}
E_{\textrm{thr}}(\lambda)=\left(\frac{1}{\sqrt{1-\frac{v^2}{c^2}}} - 1 \right)m_e c^2
\end{equation}
where $v=c/n(\lambda)$ according to Eq.~\ref{eq.vthr} and $m_e$ is the electron mass. The range of a 511--keV electron in liquid xenon is of a few mm, which is enough to produce $\sim$60 Cherenkov photons on average, as can be seen in Fig.~\ref{fig.numbCher}. The same figure shows that in the ideal case of a perfect GDE the distribution of the detected photons is almost coincident with that of produced photons. This is due to the high collection efficiency of the PETALO set-up, possible thanks to the transparency of liquid xenon to all the wavelengths involved and the almost perfect reflectivity of PTFE. 

\subsection{CRT calculation}

\begin{figure}[!bthp]
	\centering
	\includegraphics[scale=0.30]{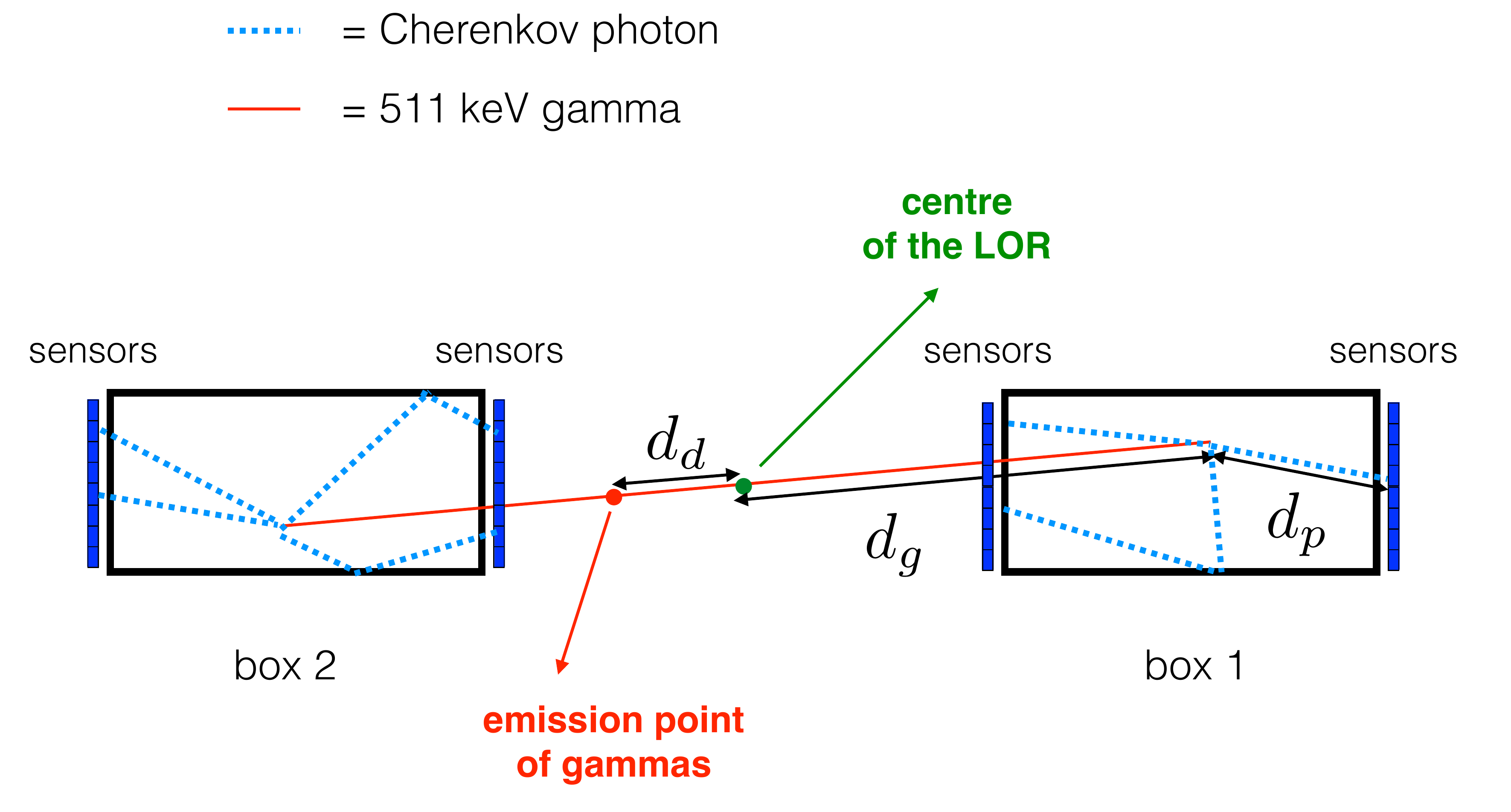}
	\caption{\label{fig.psetup} Scheme of the Monte Carlo simulation set-up, where $d_g$ is the distance between the centre of the LOR and the interaction point of the 511-keV gamma, $d_d$ is the distance between the centre of the LOR and the emission point of gammas and $d_p$ is the distance that a Cherenkov photon covers between its emission and detection point. }
\end{figure}
We denote $d_g$ as the distance from the centre of the LOR to the interaction vertex of each 511 keV gamma, $d_d$ as the displacement of the gamma emission vertex from the centre of the LOR and $d_p$ as the distance from the interaction vertex to the detection vertex (i.e., the position of the sensor), as illustrated in Fig.~\ref{fig.psetup}. If $t_1,t_2$ are the time of the first photoelectron recorded in each one of the cells, the time difference between them can be written as:
\begin{equation}
t_1 - t_2 = 2\frac{d_d}{c} + \frac{\Delta d_g}{c} + \frac{ \Delta d_p}{v}
\label{eq.first}
\end{equation}
where $v$ is the velocity of the Cherenkov photon and $c$ is the speed of light in vacuum. Therefore, the difference in time between the gamma emission vertex and the centre of the LOR can be expressed as:
\begin{equation}
\Delta t  \equiv \frac{d_d}{c}  =  \frac{1}{2}(t_1 - t_2 - \frac{\Delta d_g}{c} - \frac{\Delta d_p}{v}) 
\label{eq.CRT}
\end{equation}
The CRT, $\delta \Delta t$, is defined as the variance, expressed in FWHM, of the $\Delta t$ distribution. Therefore, the factors that affect the fluctuation of $\Delta t$ are: a) the number of detected Cherenkov photons; b) the  fluctuations in the velocity of propagation of Cherenkov photons in liquid xenon; c) the precision in the measurement of the recorded time of photoelectrons, which is driven by the time jitter of the sensors and the front-end electronics; and d) the determination of the interaction point of the 511-keV gammas, which depends on the spatial resolution of the detector.

Notice that the velocity of propagation of optical photons in Geant4 is the group velocity, which depends on the refraction index of the medium in the following way:
\begin{equation}
v = c\times \left(n(E)+\frac{\textrm{d}n}{\textrm{d}(\textrm{log}(E))}\right)^{-1}
\label{eq.velocity}
\end{equation}
where $E$ is the energy of the photon and $n$ is the refraction index of the medium. 

\section{Analysis and results}\label{sec.analysis}

\subsection{Speed of Cherenkov photons}\label{sec.speed}
\begin{figure}[!htb]
	\centering
	\includegraphics[scale=0.4]{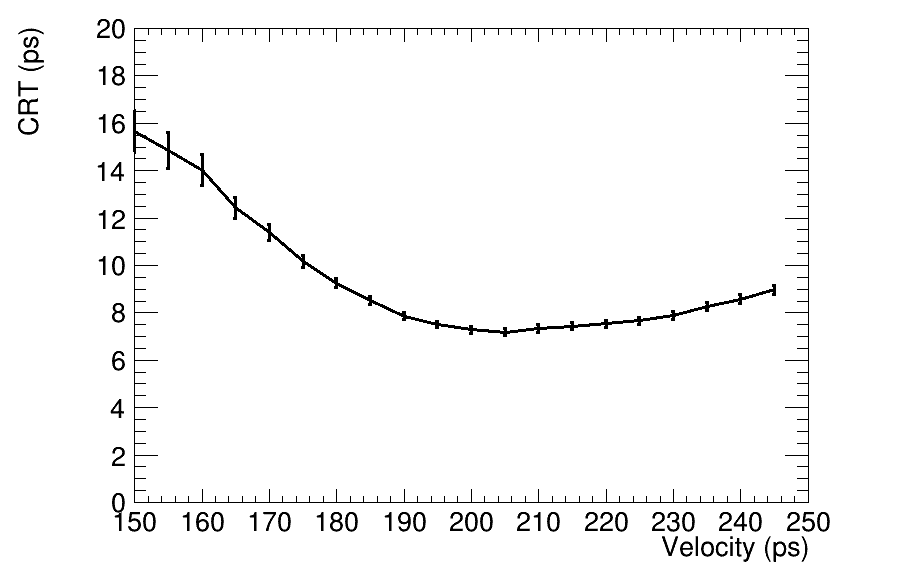}	
	\caption{\label{fig.crt_vel} Variation of the CRT as a function of the value of the optical photon speed used for the calculation. }
\end{figure}
The calculation of the CRT depends on the speed of the Cherenkov photons in liquid xenon (Eq.~\ref{eq.velocity}), which, in turn, depends on the wavelength of the photon. Since the wavelength of a detected photon is not known, it is necessary to use an average value $\bar{v}$ in the calculation of the CRT, thus introducing a fluctuation. 

Fig.~\ref{fig.crt_vel} shows the CRT as a function of  $\bar{v}$, assuming the ideal case of perfect spatial resolution and no jitter in the sensor response and in the front-end electronics. The CRT has a minimum around 200--210 mm/ns, as expected, since most of the radiation is emitted in the blue and near-UV range, where the photon speed varies very little (Fig.~\ref{fig.CherVel}-\textit{right}.) In the rest of the paper, a value of 210 mm/ns will be used for the speed of all photons, regardless of their wavelength.

\subsection{Intrinsic CRT}
%
%

\begin{figure}[!htb]
	\centering	
\includegraphics[scale=0.4]{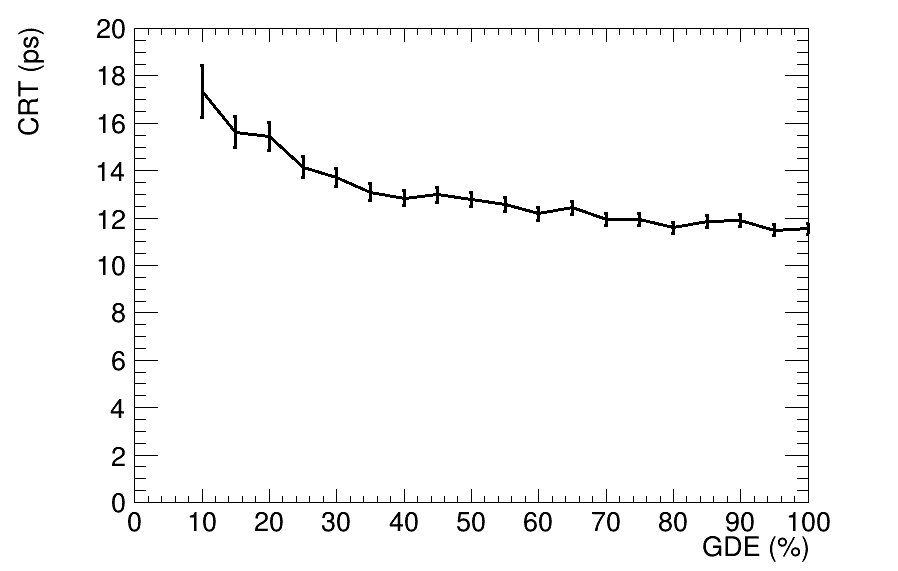}
\caption{\label{fig.allwvl} Dependence of the CRT on the sensor photodetection efficiency assuming all wavelengths are detected with the same efficiency.}
\end{figure}
Our initial calculation assumes an ideal sensor with no time jitter or fluctuations introduced by the electronics. The uncertainty in the determination of the 511--keV gamma interaction position in the cell is simulated as a gaussian fluctuation with 2--mm r.m.s., as in Ref.~\cite{Gomez-Cadenas:2016mkq}. In this case, the CRT is dominated by the GDE. Fig.~\ref{fig.allwvl} shows that the variation of the CRT with the GDE is small: using the time of the first photoelectron detected in each cell to compute the CRT, a perfect sensor with 100\% GDE exhibits a CRT of 12 ps, while a sensor with a GDE of 10\%, shows a CRT of some 17 ps.  This result demonstrates that, in spite of the low average number of detected photons in the case of low GDEs, the CRT is not affected sizeably, since it depends only on the arrival time of the first detected pair of photons.

\begin{figure}[!htb]
	\centering	
	\includegraphics[scale=0.4]{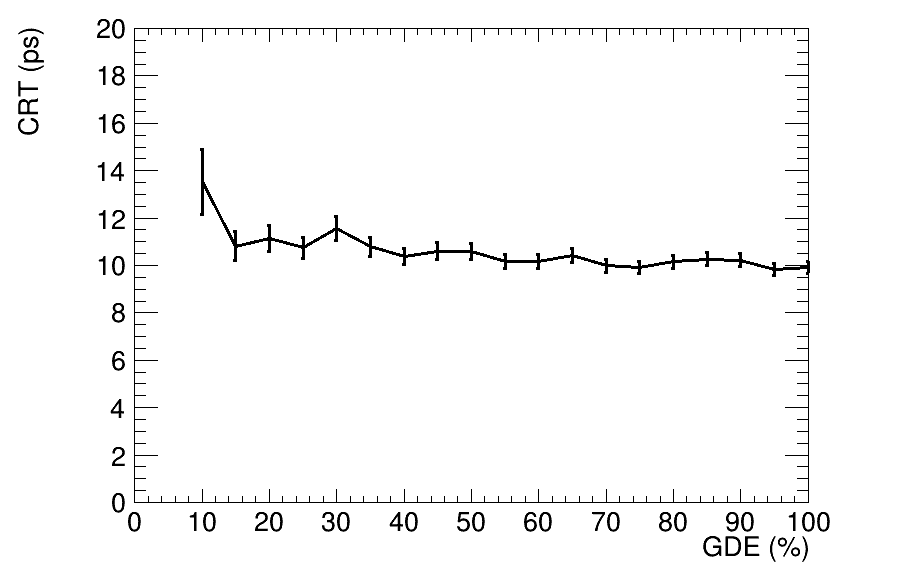}
	\caption{\label{fig.minwvl} Dependence of the CRT on the PDE with a threshold of 300 nm on the detected wavelengths.}
\end{figure}


On the other hand, Fig.~\ref{fig.CherVel}-\textit{right} shows that the speed of photons in liquid xenon varies very little for wavelengths above 300 nm. It follows that the CRT can improve using sensors with a detection threshold above 300 nm (e.g, detectors sensitive to the near UV and blue light), provided that the number of detected photons is high enough. 
Fig.~\ref{fig.minwvl} shows that, indeed,  limiting the sensitivity of the sensors down to soft UV wavelengths improves the CRT to some 10 ps almost independent of GDE. As shown in Table \ref{tab.CherProp}, the number of emitted Cherenkov photons that survive this cut is around 15--20, large enough to ensure that the CRT is not spoiled. This result has very relevant implications for our study, since it shows that: a) an intrinsic CRT of near 10 ps can be reached in a LXe detector; b) the required sensors do not need to be sensitive to hard VUV light (as is the case to detect scintillation light, where 170-nm sensitive SiPMs must be used for optimal results, as discussed in Ref.~\cite{Gomez-Cadenas:2016mkq}); and c)  it is possible to decouple the detection of scintillation light and Cherenkov light (which can be done, for example, using micro-channel plate PMTs sensitive to the near UV and optical spectrum). This decoupling allows one to optimize separately sensors dedicated to energy and position measurement and those dedicated to time measurement.


\subsection{Effect of the sensor and front-end electronics jitter}

\begin{figure}[!htb]
	\centering
	\includegraphics[scale=0.4]{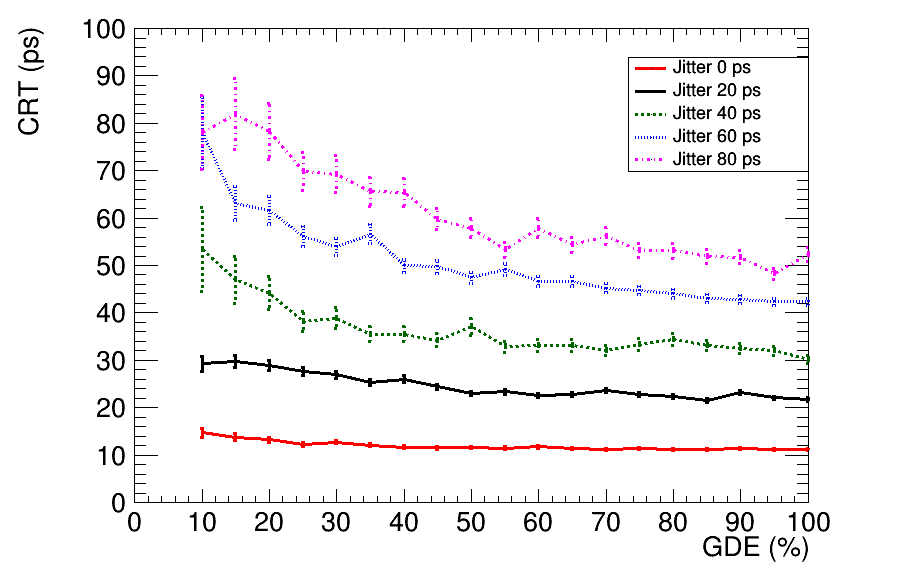}
	\caption{\label{fig.jitter_min} Dependence of the CRT on the total jitter (sensor + electronics) with a threshold of 300 nm on the detected wavelengths.}
\end{figure}

Given that the intrinsic CRT achievable with Cherenkov light in LXe approaches 10 ps, the obvious requirement for the sensors and associated electronics is to achieve a time uncertainty of the same order. With a time jitter of around 80 ps current SiPMs are far from satisfying this requirement. On the other hand, state-of-the-art fast electronics introduce a time fluctuation in the vicinity of 30 ps \cite{petsys}. When combined with very fast sensors such as single photon avalanche diodes or micro-channel plates photomultipliers, featuring time jitters of about the same order (see, for instance, Ref.~\cite{CastilloGarcia:2012} for the latter sensors), it appears that an overall time uncertainty of some 40 ps may be possible with today's technology.  

To quantify the effect, we have simulated gaussian noise (corresponding to the combined time jitter of sensors and front-end electronics) for sensors with a minimum detection wavelength of 300 nm. The results, shown in Fig.~\ref{fig.jitter_min} as a function of the GDE, show that the worst case (time jitter of 80 ps corresponding to SiPMs) yields a CRT between 50 and 80 ps (which is still better than the state-of-the-art),  while jitters around 40 ps, which appear reachable with fast detectors and electronics, result in a CRT between 30 and 55 ps, depending on the GDE. In the case of micro-channel plates, their quantum efficiency nowadays reaches 20--25$\%$ \cite{qemcp}.
%
%

%



\section{Summary and outlook}\label{sec.concl}
In this work, we demonstrate that liquid xenon, and in particular a detector along the lines of the recently proposed PETALO scanner, can effectively use Cherenkov light to provide an extraordinary CRT. The intrinsic CRT of LXe (using detectors sensitive to near UV and blue light) approaches 10 ps. In a PETALO cell, designed to cover two of the faces (entry and exit along the line of flight of the gammas) with VUV-sensitive SiPMs, one can cover the up to four additional faces with fast detectors (e.g., single photon avalanche diodes or micro-channel plates photomultipliers) sensitive to near-UV and blue light. While the CRT achieved reading scintillation light with SiPMs can be as good as 70 ps, the corresponding Cherenkov CRT may be up to a factor two better, and the combined CRT may approach 30 ps for sufficiently fast sensors and electronics. Thus, the PETALO technology may truly result in a break-through for TOF-PET scanners.

\acknowledgments
The authors acknowledge  support from the following agencies and institutions: GVA under grant PROMETEO/2016/120; the European Research Council (ERC) under the Advanced Grant 339787-NEXT; the Ministerio de Econom\'ia y Competitividad of Spain under grants FIS2014-53371-C04 and the Severo Ochoa Program SEV- 2014-0398.

\bibliography{biblio}

\end{document}